\documentclass[prd,superscriptaddress,showpacs,twocolumn]{revtex4-1}
\usepackage{graphicx}
\usepackage{units}
\usepackage{multirow}
\usepackage{bigstrut}
\usepackage{overpic}
\usepackage{array}
\usepackage{color}
\usepackage{amsmath}
\usepackage{xspace}
\usepackage{color}

\RequirePackage{lineno}
\newcommand{\gevcc}{\,\unit{GeV}/c^2}

\newcommand{\dEdx}{\ensuremath{\mathrm{d}E/\mathrm{d}x}\xspace}

\begin{document}

%\linenumbers

\newcommand{\xiaohao}{\fontsize{9pt}\baselineskip\linespread{0.5}\selectfont}
\newcommand{\dtoklenu}{D^{+} \rightarrow K^0_{L} e^{+}{\nu}_{e}}
\newcommand{\dtopipienu}{D^{+} \rightarrow K^0_{S}(\rightarrow\pi^{0}\pi^{0}) e^{+}{\nu}_{e}}
\newcommand{\dtopienu}{D^{+} \rightarrow \pi^0 e^{+}{\nu}_{e}}
\newcommand{\klenu}{K^0_{L} e^{+}{\nu}_{e}}
\newcommand{\pienu}{\pi^0 e^{+}{\nu}_{e}}
\newcommand{\genu}{ \gamma e^{+}{\nu}_{e}}
\newcommand{\de}{\Delta E}
\newcommand{\lqcd}{\Lambda_{\rm{QCD}}}
\newcommand{\ra}{$\rightarrow$}
\newcommand{\dtoenu}{D^{+} \rightarrow \gamma e^{+}{\nu}_{e}}
\newcommand{\dtolnu}{D^{+} \rightarrow \gamma l^{+}{\nu}_{l}}
\newcommand{\dkpipi}{$D^{-} \rightarrow K^{+}\pi^{-}\pi^{-}$}
\newcommand{\dkpipipi}{$D^{-} \rightarrow K^{+}\pi^{-}\pi^{-}\pi^{0}$}
\newcommand{\dkspi}{$D^{-} \rightarrow K_{S}^{0}\pi^{-}$}
\newcommand{\dkspipi}{$D^{-} \rightarrow K_{S}^{0}\pi^{-}\pi^{0}$}
\newcommand{\dkspipipi}{$D^{-} \rightarrow K_{S}^{0}\pi^{+}\pi^{-}\pi^{-}$}
\newcommand{\dkkpi}{$D^{-} \rightarrow K^{+}K^{-}\pi^{-}$}
\newcommand{\kpipi}{$K^{+}\pi^{-}\pi^{-}$}
\newcommand{\kpipipi}{$K^{+}\pi^{-}\pi^{-}\pi^{0}$}
\newcommand{\kspi}{$K_{S}^{0}\pi^{-}$}
\newcommand{\kspipi}{$K_{S}^{0}\pi^{-}\pi^{0}$}
\newcommand{\kspipipi}{$K_{S}^{0}\pi^{+}\pi^{-}\pi^{-}$}
\newcommand{\kkpi}{$K^{+}K^{-}\pi^{-}$}
\newcommand{\dd} {$D^{0}$$\bar{D}^{0}$}
\newcommand{\D}{${D^{0}}$}
\newcommand{\Dbar} {$\bar{D}^{0}$}
\newcommand{\Ks}{K_{S}^{0}}
\newcommand{\Kl}{K_{L}^{0}}
\newcommand{\mbc}{M_{\rm BC}}
\newcommand{\umiss}{U_{\rm miss}}
\newcommand{\emiss}{E_{\rm miss}}
\newcommand{\pmiss}{\vec{\bf{p}}_{\rm miss}}
\newcommand{\mmisssquare}{M^{2}_{\rm miss}}
\newcommand{\nst}{N_{\rm ST}}
\newcommand{\nsig}{N_{\rm sig}}
\newcommand{\br}{\mathcal{B}}

%Title of paper
\title{Search for the Radiative Leptonic Decay $D^{+}\to \gamma e^{+} {\nu}_{e}$}

%\input{authors_jul2015}
%\author{
%\author{Author list}
%\begin{small}
%\begin{center}
\author{%
M.~Ablikim$^{1}$, M.~N.~Achasov$^{9,e}$, S. ~Ahmed$^{14}$,
X.~C.~Ai$^{1}$, O.~Albayrak$^{5}$, M.~Albrecht$^{4}$,
D.~J.~Ambrose$^{44}$, A.~Amoroso$^{49A,49C}$, F.~F.~An$^{1}$,
Q.~An$^{46,a}$
% ...
\\
\vspace{0.2cm}
(BESIII Collaboration)\\
\vspace{0.2cm} {\it
$^{1}$ Institute of High Energy Physics, Beijing 100049, People's Republic of China\\
% ...
$^{4}$ Bochum Ruhr-University, D-44780 Bochum, Germany\\
$^{5}$ Carnegie Mellon University, Pittsburgh, Pennsylvania 15213, USA\\
% ...
$^{9}$ G.I. Budker Institute of Nuclear Physics SB RAS (BINP), Novosibirsk 630090, Russia\\
% ...
$^{14}$ Helmholtz Institute Mainz, Johann-Joachim-Becher-Weg 45, D-55099 Mainz, Germany\\
% ...
$^{44}$ University of Rochester, Rochester, New York 14627, USA\\
% ...
$^{46}$ University of Science and Technology of China, Hefei 230026, People's Republic of China\\
% ...
$^{49}$ (A)University of Turin, I-10125, Turin, Italy; (B)University of Eastern Piedmont, I-15121, Alessandria, Italy; (C)INFN, I-10125, Turin, Italy\\
% ...
$^{a}$ Also at State Key Laboratory of Particle Detection and Electronics, Beijing 100049, Hefei 230026, People's Republic of China\\
% ...
$^{e}$ Also at the Novosibirsk State University, Novosibirsk, 630090, Russia\\
% ...
}%\end{center}
%\vspace{0.4cm}
}
%}

\date{\today}

\begin{abstract}

Using an electron-positron collision data sample of 2.93 fb$^{-1}$ collected at a center-of-mass energy of $\sqrt{s}=3.773$ GeV with the BESIII detector, we present the first search for the radiative leptonic decay $D^{+} \rightarrow \gamma e^{+}{\nu}_{e}$.
The analysis is performed with a double tag method.
We do not observe a significant $D^{+} \rightarrow \gamma e^{+}{\nu}_{e}$ signal, and obtain an upper limit on the  branching fraction of  $D^{+} \rightarrow \gamma e^{+}{\nu}_{e}$ decay with the energy of radiative photon larger than 10 MeV of $3.0\times10^{-5}$ at the 90\% confidence level.

\end{abstract}

% insert suggested PACS numbers in braces on next line

%\keywords{radiative leptonic decay, BESIII}

%\maketitle must follow title, authors, abstract, \pacs, and \keywords
\maketitle

\section{Introduction}
In contrast to the purely leptonic decay, the radiative leptonic decay of 
the charged charmed meson, $\dtoenu$, is not subject to the helicity 
suppression rule due to the presence of a radiative photon.  
With no final-state hadron, treatment of the non-perturbative strong 
interaction effects in theoretical calculations is relatively simple.

The radiative leptonic decays of heavy mesons have been studied with various models~\cite{Gurdman,Geng:2000if, lv,Yang1}.
Within the perturbative quantum chromodynamics (pQCD) approach, the branching fraction of $D^{+} \rightarrow \gamma e^{+}{\nu}_{e}$ decay is predicted to be 
of order $10^{-4}$~\cite{Gurdman}.
Much smaller branching fractions, of order $10^{-6}$, are obtained within the light front quark model~\cite{Geng:2000if} and the non-relativistic constituent quark model~\cite{lv}.
In Ref.~\cite{Yang1},  the long-distance contribution is considered via 
the vector meson dominance (VMD) model and it is found that the decay rate 
may be enhanced significantly.  
To deal with non-perturbative effects, it is important to separate the 
hard and soft physics, typically with an approach known as factorization.
Many approaches to factorization of the radiative leptonic decays of heavy mesons have been proposed~\cite{factorization1,factorization2,factorization3,factorization4,factorization5,MaJP}.
In recent papers~\cite{Yang,Yang2}, factorization is extended to consider the
first-order corrections in the strong coupling constant $\alpha_s$ and 
the heavy quark mass; the branching fraction of $D^{+} \rightarrow \gamma e^{+}{\nu}_{e}$ decay is predicted to be of order $10^{-5}$.

In this paper, we present the first search for the decay $\dtoenu$, based on a data sample of 2.93~$\rm fb^{-1}$~\cite{Lum1,Lum2} collected with the BESIII detector at a center-of-mass energy $\sqrt{s} = \rm 3.773\,GeV$.
No obvious signal is observed, and an upper limit on the branching fraction of  $\dtoenu$ decay is set at the 90\% confidence level (C.L.).
In this paper, charge conjugate modes are always implied.  

\section{The BESIII detector and data set}
The BESIII detector is a general purpose spectrometer with a geometrical acceptance of 93\% of 4$\pi$.
It consists of a main drift chamber (MDC) for measuring the momentum  and specific ionization of charged particles in a 1\,T solenoidal magnetic field, a time of flight (TOF) system to perform particle identification, and a CsI(Tl) electromagnetic calorimeter (EMC) for measurement of deposited shower energies.
These components are surrounded by a multi-layer resistive plate counter 
system, which is designed to identify the muons.  A detailed description 
of the BESIII detector can be found in Ref.~\cite{BESIII}.

High-statistics Monte Carlo (MC) simulated data samples are used to determine the detection efficiency and to estimate potential background contamination.
A {\sc geant4}-based~\cite{Geant4} MC simulation program is used to simulate the interactions of particles in the spectrometer and the detector response.
For the production of $\psi(3770)$, {\sc kkmc}~\cite{kkmc} is used; it 
includes the effects of beam energy spread and initial-state radiation (ISR).
The known decay modes are generated using {\sc evtgen}~\cite{eventgen,eventgen2} according to branching fractions from the Particle Data Group (PDG)~\cite{PDG2012}, and the remaining unknown decay modes are simulated by {\sc lundcharm}~\cite{Lund}.
Final-state radiation (FSR) of charged tracks is incorporated with {\sc photos}~\cite{photons}.
In modeling the signal events, the approach of Ref.~\cite{Yang} is adopted, where first-order effects in the strong coupling constant $\alpha_s$ and the heavy quark mass are considered.  The minimum energy of the radiative photon is set at 10\,MeV to avoid the infrared divergence for soft photons.  
For $\dtopienu$ decay, which is an important background, an exclusive MC 
sample is generated by adopting the associated form-factor model and 
parameters in Ref.~\cite{Besson:2009uv}.  

\section{$\dtoenu$ Data Analysis }
The analysis uses a double-tag (DT) technique~\cite{MarkIII} 
which exploits the exclusive $D\bar{D}$ final states produced near 
threshold in $e^+e^-$ experiments. 
This technique allows one to measure absolute decay branching fractions 
of $D^+$ mesons independent of any direct knowledge of the total number 
of $D^+D^-$ events.  
In this analysis, the $D^-$ candidates, so-called single-tag (ST) 
events, are reconstructed through six specific hadronic decay modes 
\kpipi, \kpipipi, \kspi, \kspipi, \kspipipi and \kkpi.
The signal $\dtoenu$ is then searched for among the remaining tracks and 
showers recoiling against the ST $D^-$ candidates; such signal candidate 
events are denoted as double-tag (DT) events.  
The absolute branching fraction, $\br(\dtoenu)$, can be obtained from the ratio of the DT yields and the ST yields,
\begin{equation}
\br(\dtoenu) = \frac{N_{\rm DT}}{\sum_{i}N^{i}_{\rm ST}\varepsilon^{i}_{\rm DT}/\varepsilon^{i}_{\rm ST}},
\label{eq:brsig}
\end{equation}
where $N_{\rm DT}$ is the sum of signals yields for all tag modes and $N^i_{ST}$, $\varepsilon^{i}_{\rm DT}$ and $\varepsilon^{i}_{\rm ST}$ are the ST yields and the detection efficiencies of DT and ST for ST mode $i$, respectively.
With this approach, the systematic uncertainties in the ST selection 
reconstruction are largely canceled in the branching fraction measurement.

\subsection{Single-Tag event selection and yields}
For each charged track, we require the polar angle $\theta$ in the MDC to satisfy $|\cos\theta|<0.93$ and the point of the closest approach to the interaction point (IP) of the $e^+e^-$ beams to be within 1\,cm in the plane perpendicular to the beam ($V_r$) and within $\pm$10\,cm along the beam axis ($V_{z}$).
Particle identification (PID) for charged tracks is accomplished by combining 
the information on the measured ionization energy loss ($\dEdx$) in the MDC 
and the flight time in the TOF into a PID likelihood, $\mathcal{L}(h)$, for 
each hadron hypothesis $h=K$ or $h=\pi$.  
The $\pi$ $(K)$ candidates are required to satisfy $\mathcal{L}(\pi)>\mathcal{L}(K)$ ($\mathcal{L}(K)>\mathcal{L}(\pi)$).

The $K^0_S$ candidates are reconstructed from combinations of two tracks with opposite charge which satisfy $|\cos\theta|<0.93$ and $|V_{z}| < 20$\,cm, 
but with no $V_r$ and no PID requirements.  
The $K^0_S$ candidates must have an invariant mass in the range $0.487<M_{\pi^+\pi^{-}}<0.511\,\gevcc$, corresponding to three times our mass resolution.
To reject combinatorial background, we further require the decay length 
of $K^0_S$ candidates, the distance between the IP and the reconstructed secondary 
decay vertex provided by a vertex fit algorithm, to be larger than two 
standard deviations.  The momenta of $\pi^+\pi^{-}$ pairs after the vertex 
fit are used in subsequent analysis.

Those showers deposited in the EMC not associated with charged tracks 
are identified as photon candidates.
The energy deposited in the nearby TOF counters is included to improve energy resolution and detection efficiency.
The minimum deposited energy is required to be greater than 25\,MeV in the barrel region ($|\cos\theta|<0.80$), or 50\,MeV in the end caps regions ($0.84<|\cos\theta|<0.92$).
The shower time is required to be within 700\,ns after the event start time to suppress electronic noise and showers unrelated to the collision event.
The $\pi^0$ candidates are reconstructed from pairs of photons with invariant mass satisfying $0.115<M_{\gamma\gamma}<0.150\,\gevcc$; those with both photons in the EMC end caps are rejected because of poorer resolution.
The photon pairs of $\pi^0$ candidates are subject to a one-constraint (1C) kinematic fit which 
constrains their mass to the nominal $\pi^0$ mass~\cite{PDG2012}; 
the updated momenta are used in subsequent analysis.

The ST $D^-$ signals are discriminated from backgrounds based on two kinematic variables, the energy difference, $\de$, and the beam-constrained mass, $\mbc$ (encompassing energy and momentum conservation) which are  defined as:
\begin{equation}
\Delta E\equiv {E_{\rm ST}}-E_{\rm beam},
\end{equation}
\begin{equation}
M_{\rm BC} \equiv \sqrt{E_{\rm beam}^{2}/c^{4}-|{\vec{p}_{\rm ST}}|^{2}/c^{2}},
\end{equation}
where {$\vec{p}_{\rm ST}$} and {$E_{\rm ST}$} are the total momentum and energy of the {ST} {${D^-}$} candidate
in the rest frame of the $e^+e^-$ system, respectively, and $E_{\rm beam}$ is the beam energy.
The ST signals peak around zero in the $\Delta E$ distribution and around the nominal $D^-$ mass~\cite{PDG2012} in the $M_{\rm BC}$ distribution.

\begin{figure}[htbp]
\includegraphics[width=0.49\textwidth]{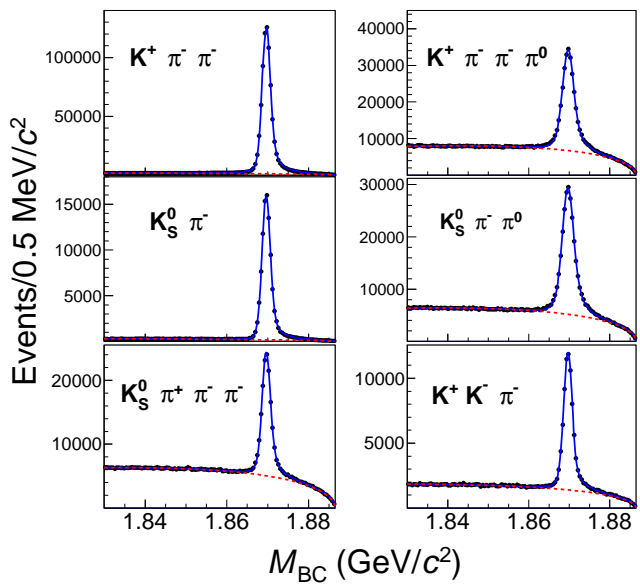}
\caption{(color online) The $M_{\rm BC}$ distributions for the six tag modes.
Dots with error bars are data. The blue solid lines show the overall fit curves and the red dashed lines are for the background contributions.
\label{fig:mbcfits}}
\end{figure}

For each ST mode, the $D^-$ candidates are reconstructed from all possible combinations of final-state particles,
and are required to have $\de$ within the regions listed in Table~\ref{table:st_yields}; these are final-state dependent and determined from data.
If multiple candidates are found, only the one with the smallest $|\de|$ is selected.
To extract the ST signal yields, we perform extended unbinned maximum likelihood fits to the $M_{\rm BC}$ distributions, as shown in Fig.~\ref{fig:mbcfits}.
In the fits, signal shapes derived from the signal MC events are convoluted 
with a Gaussian function; the free mean and width of this Gaussian 
compensate for imperfections in the beam energy calibration and differences 
in the detector resolution between data and MC simulation, respectively.  
The combinatorial background is modeled by a smooth ARGUS function~\cite{Argus}.
The signal yields  and the corresponding detection efficiencies in the region $1.8628 <M_{\rm BC}<1.8788 \gevcc$ are summarized in Table~\ref{table:st_yields}.
A study of the inclusive $D\bar{D}$ MC samples, in which both $D$ mesons 
decay inclusively, indicates that there are no significant backgrounds 
which peak in $M_{BC}$.

\begin{table}[htbp]
\caption{ Summary of the $\de$ requirements, ST yields $N_{\rm ST}^{i}$ in data and detection efficiencies $\varepsilon_{\rm ST}^{i}$. The efficiencies do not include the branching fractions of $K_S^0\to\pi^+\pi^-$ and $\pi^0\to\gamma\gamma$.  All uncertainties are statistical only.}
\label{table:st_yields}
\begin{ruledtabular}
\begin{tabular}{lcccc}
Tag Mode    &$\de$ (MeV)  &$N_{\rm ST}^{i}$&  $\varepsilon_{\rm ST}^{i}$(\%) \\
\hline
\kpipi     &$[-27, 25]$  & 801498 $\pm$ 940  & 51.57 $\pm$ 0.02 \\
\kpipipi   &$[-62, 34]$ & 242092 $\pm$ 699  & 24.37 $\pm$ 0.02 \\
\kspi      &$[-25, 25]$ &  98132 $\pm$ 328  & 54.03 $\pm$ 0.06 \\
\kspipi    &$[-73, 41]$ &  213976 $\pm$ 641  & 26.17 $\pm$ 0.02 \\
\kspipipi  &$[-33, 30]$ & 127463 $\pm$ 415  & 32.46 $\pm$ 0.04 \\
\kkpi      &$[-23, 20]$ &  70701 $\pm$ 343 & 41.83 $\pm$ 0.06 \\
\end{tabular}
\end{ruledtabular}
\end{table}

\subsection{Double-Tag event selection and yields}
We search for the signal $\dtoenu$ in the remaining charged tracks and showers recoiling against the ST $D^-$ candidates.
Exactly one good remaining charged track is required, with charge opposite to that of the ST $D^-$.
The track must be identified as an electron by combining the information 
from $\dEdx$, TOF, and the EMC.
The PID $\mathcal{L}$ is required to satisfy $\mathcal{L}(e)>0$ and  $\mathcal{L}(e)/(\mathcal{L}(e)+\mathcal{L}(\pi) + \mathcal{L}(K))>0.8$.
There must be at least one remaining 
photon to be selected as the candidate radiative photon; 
in the case of multiple candidates, the highest energy photon is used.
However, we reject events in which any pair of photons satisfies $\chi^2<20$ 
in the $\pi^0$ 1C kinematic fit.  
To improve the degraded momentum resolution of the electron due to FSR 
and bremsstrahlung, the energy of neighboring photons, presumably due to FSR, 
is added back to electron candidates.  
Specifically, photons with energy greater than 50\,MeV and within a cone of 5 degrees around the electron direction (but excluding the radiative one) are 
included.  
To suppress the background $\dtoklenu$, the radiative photon is further required to have a lateral moment~\cite{lateral}
within the range (0.0, 0.3).  
This lateral moment, which describes the shape of electromagnetic showers, is found in MC event studies to peak around 0.15 for photons but to vary broadly from 0 to 0.85 for $K^0_{L}$ candidates.

\begin{figure}[htbp]
\includegraphics[width=0.4\textwidth]{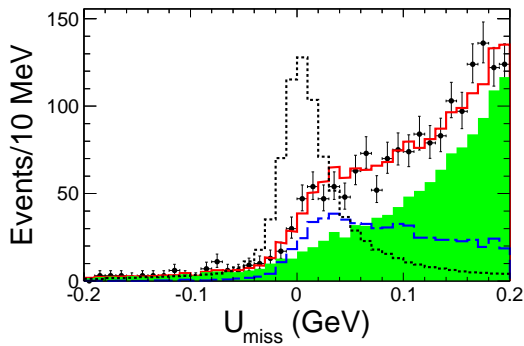}
\caption{(color online) The $\umiss$ distribution. Dots with error bars are data, the red solid-line histogram shows the overall fit curve, the blue dash-line histogram shows the background $\dtopienu$, and the green shaded histogram includes all other background.  The black dotted line shows the signal MC simulation normalized to the branching fraction $\mathcal{B}(\dtoenu) = 100 \times 10^{-5}$.
\label{fig:ufit}}
\end{figure}

In the selection of DT events, the undetected neutrino is inferred by studying the missing energy, $E_{\rm miss}$, and missing momentum, $\vec{p}_{\rm miss}$,
which are defined as
\begin{equation}
E_{\rm miss} \equiv E_{\rm beam}-E_{\gamma} - E_{e} ,
\end{equation}
and
\begin{equation}
\vec{p}_{\rm miss} \equiv -[\vec{p}_{\gamma}+\vec{p}_{e}+ \hat{p}_{\rm ST}\sqrt{E_{\rm beam}^{2}/c^2-m_{D^-}^{2}c^2}],
\end{equation}
in the rest frame of  $e^+e^-$ system.
Here, $E_{\gamma}$ ($E_{e}$) and $\vec{p}_{\gamma}$ ($\vec{p}_{ e}$) are the energy and momentum of the radiative photon (electron), respectively, and $m_{D^-}$ is the nominal mass of the $D^-$ meson~\cite{PDG2012}.
In calculating $\vec{p}_{\rm miss}$, only the direction vector of the ST $D^-$ candidate, $\hat{p}_{\rm ST}$, is used; the corresponding magnitude of momentum is fixed.
The variable $U_{\rm miss}$ is then defined as
\begin{equation}
U_{\rm miss}\equiv E_{\rm miss}-|\vec{p}_{\rm miss}|c. \label{eq:ucal}
\end{equation}
The distribution of $U_{\rm miss}$ for the surviving DT candidates is illustrated in Fig.~\ref{fig:ufit}; the $\dtoenu$ signals should peak around zero, 
as shown with the dotted curve.

By studying the MC simulation samples, the background from the semi-leptonic decay $\dtopienu$ is found to have a non-trivial shape in $U_{\rm miss}$. 
Therefore, we study the $\dtopienu$ backgrounds exclusively by selecting a control sample in data with exactly the same selection criteria for the ST events and electron candidates used in the selection of signal events.  
The $\pi^0$ candidates are reconstructed from two photons with a 1C kinematic fit constraining their mass to the $\pi^0$ nominal value and having a fit $\chi^2<20$.
We extract the yield of the control sample $\dtopienu$, $N^{\pi^0}_{\rm DT}$, by fitting the corresponding $U_{\rm miss}$ distribution.
The expected number of background $\dtopienu$ in the selection of signal $\dtoenu$, $N_{\pi^0}^{\rm exp}$, is calculated with
\begin{equation}
N_{\pi^0}^{\rm exp} =\frac{N^{\pi^0}_{\rm DT}} {\sum_{i} \frac{N_{\rm ST}^i}{\varepsilon_{\rm ST}^i} {\varepsilon^{i}_{{\rm DT}, \pi^0}}} \sum_{i} \frac{N_{\rm ST}^i}{\varepsilon_{\rm ST}^i} \varepsilon^{i,\gamma}_{{\rm DT},\pi^0},
\label{eq:cal_npi0}
\end{equation}
where $\varepsilon^{i}_{{\rm DT},\pi^0}$ is the DT efficiency of $\dtopienu$, $\varepsilon_{{\rm DT},\pi^0}^{i,\gamma}$ is the rate of mis-identifying $\dtopienu$ as $\dtoenu$ for the tag mode $i$, individually.
The values of the corresponding efficiencies are summarized in Table~\ref{table:dt_effi}.
We find $N^{\pi^0}_{\rm DT} = 3016 \pm 68$ 
and $N_{\pi^0}^{\rm exp} = 612 \pm 14$, respectively, 
where the errors are statistical only.

\begin{table}[htbp]
\caption{Summaries of the DT efficiencies of $\dtoenu$ ($\varepsilon^{i}_{\rm DT}$) and $\dtopienu$ ({$\varepsilon^i_{\rm DT,\pi^0}$}), and the rates of mis-identifying $\dtopienu$ as $\dtoenu$ ($\varepsilon_{{\rm DT}, \pi^0}^{i,\gamma}$), where the branching fraction of $K_S^0\to\pi^+\pi^-$ and $\pi^0\to\gamma\gamma$ are not included.
The uncertainties are MC statistical only.}
\label{table:dt_effi}
\begin{ruledtabular}
\begin{tabular}{lccc}
Tag Mode    & $\varepsilon^{i}_{\rm DT}$(\%) & {$\varepsilon^i_{\rm DT,\pi^0}$} (\%)& $\varepsilon_{\pi^0}^{i,\gamma}$  (\%) \\
\hline
\kpipi     & 27.09 $\pm$ 0.11 & 27.93 $\pm$ 0.14 & 5.32 $\pm$ 0.07 \\
\kpipipi   & 14.28 $\pm$ 0.08 & 13.79 $\pm$ 0.11 & 3.05 $\pm$ 0.05 \\
\kspi      & 28.97 $\pm$ 0.10 & 30.23 $\pm$ 0.14 & 5.87 $\pm$ 0.07\\
\kspipi    & 15.62 $\pm$ 0.08  & 15.17 $\pm$ 0.11 & 3.29 $\pm$ 0.06\\
\kspipipi  & 17.86 $\pm$ 0.09 & 17.55 $\pm$ 0.12 & 3.72 $\pm$ 0.06 \\
\kkpi      & 21.12 $\pm$ 0.10 & 22.28 $\pm$ 0.13 & 4.19 $\pm$ 0.06\\
\end{tabular}
\end{ruledtabular}
\end{table}

An extended unbinned maximum likelihood fit is performed on the final $\umiss$ distribution as shown in Fig.~\ref{fig:ufit}.
The signal shape is derived from the simulated $\dtoenu$ events convoluted with a Gaussian function to compensate for resolution differences between data and MC simulation.
The parameters of this Gaussian smearing function are extracted according to the discrepancy in resolution between data and MC simulation in the control sample $\dtopienu$, and are fixed in the fit.
The shape of the background $\dtopienu$ is extracted from the simulated $\dtopienu$ sample, and is normalized to $N_{\pi^0}^{\rm exp}$.  
For the other background components, the shape from the inclusive MC sample (excluding the contribution from $\dtopienu$) is adopted and the yield is determined in the fit.
We obtain a signal yield of $N_{\rm DT}=-21 \pm 23$, and the resulting branching fraction is $\mathcal{B}(\dtoenu) = (-2.5 \pm 2.7) \times10^{-5}$, where the uncertainties are statistical only.
Since no obvious signal is observed, an upper limit at the 90\% C.L. on the branching fraction of $\dtoenu$ will be set below after taking into account the effects of statistical and systematic uncertainties.

\section{Systematic Uncertainties}
The systematic uncertainties in the selection of the ST candidates are assumed 
to largely cancel, with any residual effects being negligible.  
Other systematic uncertainties, related to the detection efficiencies, are summarized in Table~\ref{table:sys}.
To evaluate the systematic uncertainty related to the model of the decay dynamics, an alternative signal MC sample based on the single pole model~\cite{Gurdman,Yang} is produced, and the resultant difference in the detection efficiency with respect to the nominal value, 3.5\%, is assigned as the systematic uncertainty.
The uncertainties of electron tracking and PID are estimated to be 0.5\% and 0.5\%, respectively, by studying a control sample of radiative Bhabha scattering events.
The uncertainty in photon reconstruction is assigned as 1.0\%, based on a study of double-tagged $D^{0} \rightarrow K_{S}\pi^0$ events~\cite{gammapaper}.
The uncertainty related with the lateral moment requirement for the photon is estimated to be 4.4\% by studying a photon control sample from radiative Bhabha scattering events.
The quadratic sum of the above systematic uncertainties, related to detection efficiency, is 5.8\%.

The systematic uncertainty associated with the estimated number of background $\dtopienu$ events includes a statistical uncertainty on the size of the DT control sample ($\dtopienu$) of 2.3\%, and relative uncertainties on the detection efficiency relative to signal, of 1.0\% for the $\pi^0$ 1C kinematic fit, 
and 1.0\% for the extra photon with respect to the signal.  
Adding in quadrature, the total uncertainty of the background $\dtopienu$ rate is 2.7\%.   
Note this value is not the direct fractional change in the branching fraction of $\dtoenu$, it is the fluctuation of background $\dtopienu$ and will be considered along with other effects from the fit procedure.  

Various sources of systematic uncertainties in the fit procedure are considered:
({\it a}) fits are redone with the fitting range being as $(-0.15,0.25)$\,GeV or $(-0.20,0.25)$\,GeV;
({\it b}) the mean and width of the smearing Gaussian function for the signal shape are varied according to the corresponding uncertainties obtained from the control sample $\dtopienu$;
({\it c}) the number of the background $\dtopienu$ is varied according its uncertainty (2.7\%);
({\it d}) the shape derived from the inclusive MC sample is replaced by a second order polynomial function to describe the other backgrounds excluding $\dtopienu$.  All of these fitting procedure effects are accounted for within the upper limit evaluation described next.  

\begin{table}[htbp]
\caption{Systematic uncertainties related to detection efficiencies in the branching fraction measurement.}
\label{table:sys}
\begin{ruledtabular}
\begin{tabular}{lc}
Source  & Relative uncertainty (\%)  \\
\hline
Signal MC model &  3.5 \\
{$e^+$ t}racking &  0.5    \\
$e^+$ PID  &  0.5    \\
$\gamma$ reconstruction  & 1.0   \\
Lateral moment &  4.4    \\
 $\pi^{0} e^+ \nu_{e} $ {backgrounds} &  2.7\footnote{Note, this value is a fractional change in the $\pienu$ rate, not in the branching fraction of $\dtoenu$.}    \\
\end{tabular}
\end{ruledtabular}
\end{table}

\section{The upper limit on branching fraction}
\begin{figure}[htbp]
\includegraphics[width=0.4\textwidth]{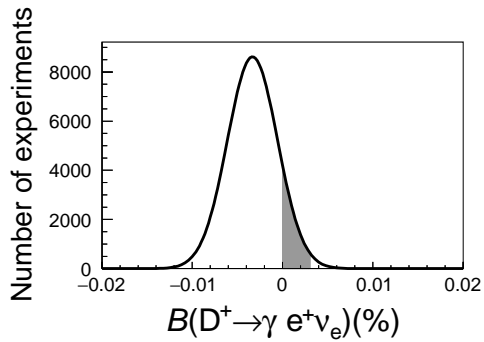}
\caption{Distribution of the accumulated branching fraction based on toy MC samples generated according to the data. The shaded region represents 90\% of the physical region.
\label{fig:BR}}
\end{figure}

To set the upper limit on the decay branching fraction $\br(\dtoenu)$, we follow the method in Refs.~\cite{BaBar,gammapaper} which takes into account the effects of both systematic and statistical uncertainties.
We obtain a smooth probability density function (PDF) from the data sample using the kernel estimation method~\cite{KEM}.
A large number of toy MC samples are generated according to the smooth PDF, while the number of events in each MC sample is allowed to fluctuate with a Poisson distribution according to the yield found in the fit to the data sample.
The same fit procedure used for data is applied to each toy MC sample, while randomly making systematic variations in the fit procedure, as described in the previous section.
In the calculation of the branching fraction $\br(\dtoenu)$ for the toy MC sample, the DT efficiencies are varied randomly according to the detection efficiency uncertainties (5.8\%), and the ST yields and the corresponding efficiencies are varied randomly according to the statistical uncertainty due to the size of data and MC samples.
The resultant distribution of $\br(\dtoenu)$ for all toy MC samples is shown in Fig~\ref{fig:BR}.
By integrating up to 90\% of the area in the physical region $\br(\dtoenu)\geq 0$, we obtain an upper limit at the 90\% C.L. for the branching fraction as $\br(\dtoenu)<3.0\times10^{-5}$.

\section{Summary}
In summary, we present the first search for the radiative leptonic decay $\dtoenu$ in the charm sector based on a DT method
using a data sample of 2.93$\rm\,fb^{-1}$ collected with the BESIII detector at a center-of-mass energy $\sqrt{s}=3.773$\,GeV.
No significant $\dtoenu$ signal is observed.
With a 10\,MeV cutoff on the radiative photon energy, the upper limit of the decay branching fraction for $\dtoenu$ is $\br(\dtoenu)< 3.0\times10^{-5}$ 
at the 90\% C.L.  
The result approaches the theoretical predictions in Refs.~\cite{Yang,Yang2}; 
more data may help to discriminate among the full suite of theoretical models.

\begin{acknowledgments}
The BESIII collaboration thanks the staff of BEPCII and the IHEP computing center for their strong support. This work is supported in part by National Key Basic Research Program of China under Contract No. 2015CB856700; National Natural Science Foundation of China (NSFC) under Contracts Nos. 11125525, 11235011, 11275266, 11322544, 11335008, 11425524, 11635010; the Chinese Academy of Sciences (CAS) Large-Scale Scientific Facility Program; the CAS Center for Excellence in Particle Physics (CCEPP); the Collaborative Innovation Center for Particles and Interactions (CICPI); Joint Large-Scale Scientific Facility Funds of the NSFC and CAS under Contracts Nos. U1232201, U1332201, U1532257, U1532258; CAS under Contracts Nos. KJCX2-YW-N29, KJCX2-YW-N45, QYZDJ-SSW-SLH003; 100 Talents Program of CAS; National 1000 Talents Program of China; INPAC and Shanghai Key Laboratory for Particle Physics and Cosmology; German Research Foundation DFG under Contracts Nos. Collaborative Research Center CRC 1044, FOR 2359; Istituto Nazionale di Fisica Nucleare, Italy; Koninklijke Nederlandse Akademie van Wetenschappen (KNAW) under Contract No. 530-4CDP03; Ministry of Development of Turkey under Contract No. DPT2006K-120470; The Swedish Resarch Council; U. S. Department of Energy under Contracts Nos. DE-FG02-05ER41374, DE-SC-0010504, DE-SC-0010118, DE-SC-0012069; U.S. National Science Foundation; University of Groningen (RuG) and the Helmholtzzentrum fuer Schwerionenforschung GmbH (GSI), Darmstadt; WCU Program of National Research Foundation of Korea under Contract No. R32-2008-000-10155-0.
\end{acknowledgments}

% Create the reference section using BibTeX:
%\bibliography{basename of .bib file}
\bibliography{mybib}
\end{document}